\documentclass[journal]{IEEEtran}
\ifCLASSINFOpdf
\else
\fi
\usepackage{booktabs}
\usepackage{multirow}
\usepackage{graphicx}
\usepackage{subfigure}
\usepackage{amsmath}
\usepackage{amssymb}
\usepackage{cite}
\usepackage{threeparttable}
\usepackage{adjustbox}
\usepackage{color}

\begin{document}

\title{Scale-aware Super-resolution Network with Dual Affinity Learning for Lesion Segmentation from Medical Images}


\author{Yanwen~Li*, %(Imsight AI Research Lab) \<liyanwen@imsightmed.com\>
Luyang~Luo*,~\IEEEmembership{Member,~IEEE,} %(The Chinese University of Hong Kong) \<lyluo@cse.cuhk.edu.hk\>
Huangjing~Lin, %(The Chinese University of Hong Kong) \<hjlin@cse.cuhk.edu.hk\>
Pheng-Ann Heng,~\IEEEmembership{Senior~Member,~IEEE, } %(The Chinese Univsersity of Hong Kong)<pheng@cse.cuhk.edu.hk>}
Hao Chen\#,~\IEEEmembership{Senior~Member,~IEEE} %(The Hong Kong University of Science and Technology)<jhc@cse.ust.hk>

\thanks{The work described in this paper was partially supported by the Hong Kong Innovation Fund (No. ITS/028/21FP), Shenzhen Science and Technology Innovation Committee (No. SGDX20210823103201011), and a grant from the Research Grants Council of the Hong Kong Special Administrative Region, China (Project Reference Number: T45-401/22-N).}
\thanks{*Yanwen~Li and Luyang~Luo contributed equally.}
\thanks{Yanwen~Li and Huangjing~Lin are with the Imsight AI Research Lab.(e-mails: \{liyanwen, linhuangjing\}@imsightmed.com).}
\thanks{Luyang~Luo and Hao Chen are with the Department of Computer Science and Engineering, The Hong Kong University of Science and Technology (e-mails: cseluyang@ust.hk, jhc@cse.ust.hk). Hao Chen is also with the Department of Chemical and Biological Engineering, The Hong Kong University of Science and Technology.}
\thanks{Pheng-Ann Heng is with the Department of Computer Science and Engineering, The Chinese University of Hong Kong (e-mail: pheng@cse.cuhk.edu.hk)}
\thanks{\# denotes the corresponding author.}
}

% <-this % stops a space
% \thanks{}

\markboth{Journal of \LaTeX\ Class Files}%
{Shell \MakeLowercase{\textit{et al.}}: Bare Demo of IEEEtran.cls for IEEE Journals}



\maketitle

\begin{abstract}
   Convolutional Neural Networks (CNNs) have shown remarkable progress in medical image segmentation.
   However, lesion segmentation remains a challenge to state-of-the-art CNN-based algorithms due to the variance in scales and shapes.
   \textcolor{black}{On the one hand, tiny lesions are hard to be delineated precisely from the medical images which are often of low resolutions.
   On the other hand, segmenting large-size lesions requires large receptive fields, which exacerbates the first challenge.}
   In this paper, we present a scale-aware super-resolution network to adaptively segment lesions of various sizes from the low-resolution medical images.
   Our proposed network contains dual branches to simultaneously conduct lesion mask super-resolution and lesion image super-resolution. 
   \textcolor{black}{The image super-resolution branch will provide more detailed features for the segmentation branch, i.e., the mask super-resolution branch, for fine-grained segmentation.} 
   Meanwhile, we introduce scale-aware dilated convolution blocks into the multi-task decoders to adaptively adjust the receptive fields of the convolutional kernels according to the lesion sizes.
   To guide the segmentation branch to learn from richer high-resolution features, we propose a feature affinity module and a scale affinity module to enhance the multi-task learning of the dual branches.
   On multiple challenging lesion segmentation datasets, our proposed network achieved consistent improvements compared to other state-of-the-art methods. 
   %Code will be released.%Code are available at https://github.com/poiuohke/SASR.
\end{abstract}


% \fi
\IEEEpeerreviewmaketitle

\section{Introduction}

\IEEEPARstart{L}{esions} are the areas of abnormal tissues usually caused by diseases or traumas. 
Lesion segmentation from medical images could assist doctors to detect and locate diseases, providing them with detailed and quantitative analysis for more accurate diagnosis, prescriptions, and longitudinal monitoring.
\textcolor{black}{For example, the segmentation of lung nodules from computed tomography(CT) images could offer morphological parameters which are important for malignancy assessment}.
The segmentation of skin lesions from dermoscopy images could assist dermatologists in clinical evaluation for a more accurate and efficient diagnosis. 
The segmentation of polyps could provide references for subsequent resections.
Conventional lesion segmentation methods are often based on region growing or optimum thresholding, which are usually \textcolor{black}{expert-depending, parameter-sensitive, and time-consuming.}
%usually require manual intervention, hyperparameters tuning for specific scenes, and a lot of time.
As a promising solution to alleviating the burden on the doctors, convolutional neural network-based (CNN-based) automatic segmentation methods have recently made remarkable achievements in medical image segmentation \cite{shen2017deep}. 

\begin{figure}[!t]
\begin{center}
	\includegraphics[width=0.75\columnwidth]{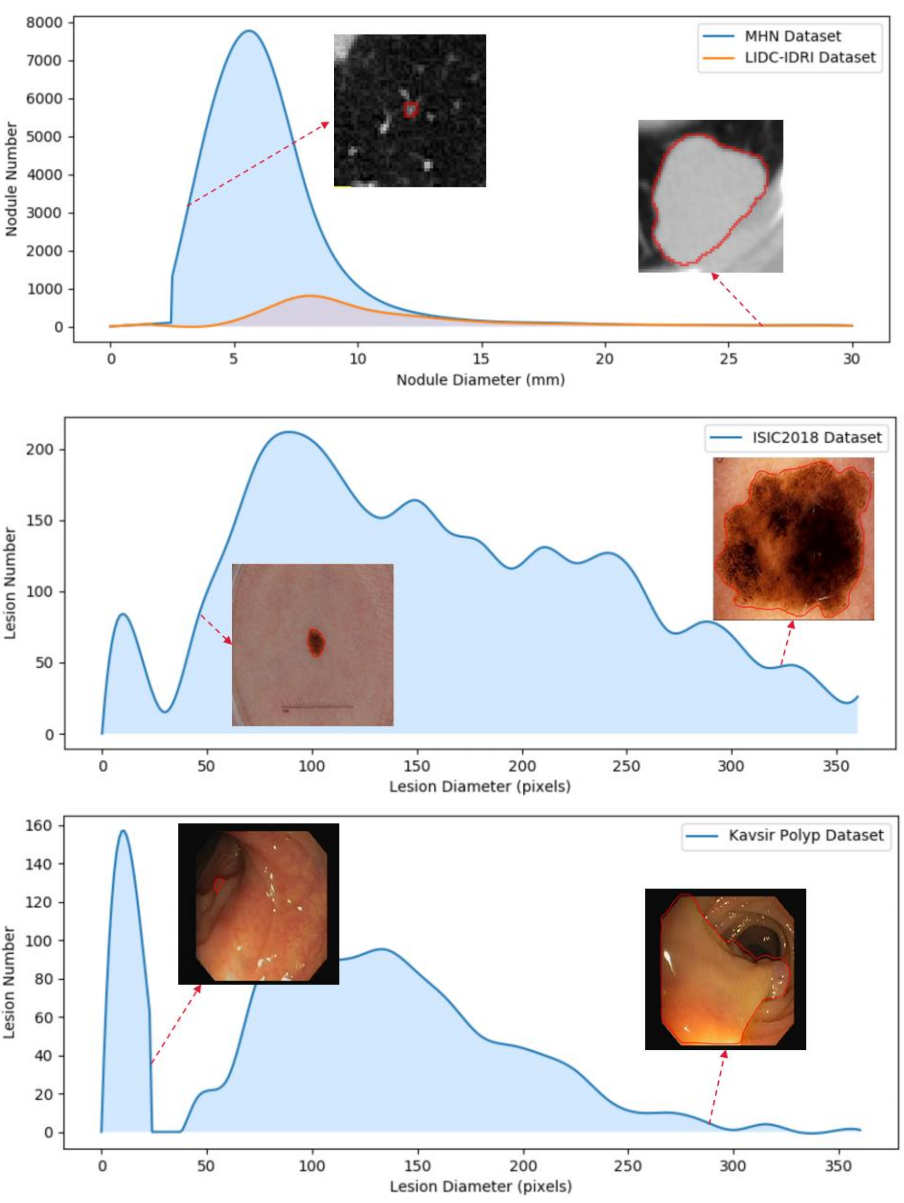}
\end{center}
\caption{The distribution of lesions with different sizes in several lesion segmentation datasets. As shown, the lesion size covers a large range and is unevenly distributed in the different datasets. The large differences between the tiny lesions and the large lesions may cause difficulties for the segmentation task.}
\label{data_spacing}
\end{figure}

However, lesions are often diverse in sizes, \textcolor{black}{posing challenges to modern CNN-based methods.}
\textcolor{black}{On the one hand, features of the tiny lesions are hard to capture due to the series of downsamplings in CNNs.}
\textcolor{black}{On the other hand, some lesions could grow as large as tens or hundreds of times of the tiny ones and the context information is hence difficult to be captured.}
Hence, the large lesions require a big receptive field to capture the complete patterns, despite that the tiny lesions would need a smaller receptive field to better retain the original small amount of information.

\textcolor{black}{Different efforts have been devoted to generate and aggregate multi-scale features to capture diversely sized targets~\cite{zhao2017pyramid,chen2017deeplab}.}
Other methods may adopt Unet-like frameworks and build skip connections among layers of different depths to generate multi-scale representations~\cite{gao2019res2net}.
These solutions more or less rely on enlarging the receptive field, and the inputs are often required to be high-resolution.
Nevertheless, medical images are often of low resolutions. 
Taking chest CT scans with 0.6 mm spacing and 512$\times$512 in resolution as an example, the tiny nodules could show only several pixels on the CT images, which makes the detailed information more illegible. 
\textcolor{black}{Thus,} achieving accurate lesion segmentation is faced with two crucial challenges raised by the diversity of the lesion sizes and the low resolution of medical images.

To tackle the mentioned dilemma, we propose a scale-aware super-resolution network with dual affinity learning for \textcolor{black}{segmenting the diversely sized lesions.}
Specifically, we utilize a dual-path structure which consists of a lesion mask super-resolution (LMSR) path and a lesion image super-resolution (LISR) path.
The dual branches simultaneously generate super-resolution reconstructions and segmentation results from the inputs, which helps recover the fine-grained information of especially the tiny lesions from the relatively low-resolution medical images.
We then introduce scale-aware dilated convolution (SDC) blocks to enable dynamic adjustment of the model's receptive field according to the lesion sizes.
We further propose a feature affinity module to enhance the feature learning of the LMSR branch with guidance from the LISR branch.
Moreover, as the different objectives of the dual branches lead to different emphases on the foregrounds and backgrounds, a scale affinity module is introduced to align the receptive fields of the two branches and help stabilize the learning process.

We have conducted extensive experiments on multiple challenging lesion segmentation datasets: a self-collected multi-hospital nodule segmentation dataset, the LIDC-IDRI nodule segmentation dataset \cite{armato2011lung}, the ISIC2018 skin lesion segmentation dataset \cite{codella2019skin}, and the multi-source polyp segmentation benchmark \cite{fan2020pranet}. We showed that our proposed method achieved state-of-the-art performance on all four datasets with consistent improvement over other competitive approaches. 

Our main contributions can be summarized as follows:
\begin{itemize}
    \item We developed a dual-path lesion segmentation structure, with lesion mask super-resolution and lesion image super-resolution branches, to improve the segmentation of tiny lesions from low-resolution medical images.
    \item We introduced scale-aware dilated convolution blocks, which promotes segmenting lesions of various sizes by dynamically adjusting the model's receptive field.
    \item We proposed dual affinity learning modules, including feature affinity learning and scale affinity learning, to enhance the learning of the LMSR branch by the guidance of richer features from the LISR path.
    \item We demonstrated with extensive experiments that our method consistently outperforms other state-of-the-art approaches on multiple challenging datasets.
\end{itemize}

The remainders of this paper are organized as follows. We review the related work in Section~\ref{related_work} and elaborate on the proposed method in Section~\ref{method}. We present experiments and results in Section~\ref{experiments} and finally draw the conclusions in Section~\ref{conclusion}.

\section{Related Works} \label{related_work}
In this section, we briefly review previous works on lesion segmentation, multi-scale image segmentation, and super-resolution methods for other vision tasks.

\subsection{Lesion Segmentation}
Lesion segmentation plays an essential role in disease diagnosis, treatment planning, and follow-up monitoring for medical image analysis.
\textcolor{black}{Unet \cite{ronneberger2015u} and its variants have shown great success in this field \cite{azad2022medical}.}
Unet introduced the skip connections into the fully convolutional encoder-decoder structure and hence largely retained the detailed information from the original input image.
Unet++ \cite{zhou2019unet++} redesigned the skip connections of Unet and enabled feature aggregation from different semantic scales.
H-DenseUnet \cite{li2018h} aggregates both the 2D features and 3D volumetric contexts for liver tumors and optimized the network through a hybrid feature fusion.
Yu et al. proposed an offline and online 3D integration framework for polyp lesion detection from colonoscopy videos \cite{yu2016integrating} and solved the limited data problem.

The fusion of the semantic information from the deeper layers with larger receptive fields and the detail information from the shallow layers were proved capable of effectively enhancing the performance of lesions \cite{fan2020pranet,dai2022ms,yu2016automated,chen2018voxresnet,wu2020automated,sarker2018slsdeep}.
For example, PraNet \cite{fan2020pranet} aggregated the features from multi-level layers and utilized the reverse attention module to establish the cooperation mechanism between the lesion areas and boundaries for polyp segmentation.
FCRN \cite{yu2016automated} and VoxResNet \cite{chen2018voxresnet} adopted residual learning to build a deep CNN and incorporated a multi-scale contextual information integration scheme by generating several prediction maps from different levels of features for skin lesion or brain tissues segmentation.
There were also works that fused multi-scale features learned from multi-resolution inputs~\cite{kamnitsas2017efficient, dou2016multilevel,Lin_2017_CVPR}.

\textcolor{black}{However, these methods could be infeasible for segmenting tiny lesions which might show only a few pixels on the image.
Our method, on the contrary, addressed the challenge raised by low-resolution medical images by the proposed dual super-resolution structure.}

\subsection{Multi-scale Image Segmentation}
Multi-scale image segmentation has been studied quite a lot in the literature.
Multiple dilated convolutional kernels could be used to extract multi-scale features with different receptive fields \cite{yu2015multi,chen2017deeplab} for further fusion and segmentation.
PSPNet \cite{zhao2017pyramid} fused multiple pooled features, which also enabled multi-scale receptive fields aggregation for the network.
DCAN \cite{chen2016dcan} utilized multi-level contextual feature representations and investigated the complementary information as objects and contours under a multi-task learning framework.

There were also methods using context-aware filters and depth-wise convolution to capture multi-scale feature representation \cite{he2019dynamic} and then concatenate the features for final segmentation.
Different multi-scale feature aggregation strategies were also proposed.
Multi-scale context intertwining (MSCI) \cite{lin2018multi} exchanged information between adjacent scales bidirectionally and recurrently via LSTM chains.
Context contrasted local model \cite{ding2018context} applied a gated sum to select appropriate scale features for each position.
Another possible solution was using deep attention with auxiliary losses to find out the features to be used from a more appropriate layer \cite{sinha2020multi}.

\textcolor{black}{Previous methods often treated the multi-scale features equally and fused them with the same weights,} lacking the ability to adapt the receptive fields according to the target size.
To address such a limitation, we propose the scale-aware dilated convolution blocks to dynamically adjust the receptive fields according to the input lesion size.

\subsection{Super Resolution for Vision Tasks}
Single image super-resolution (SISR) aims to recover high-resolution (HR) images from low-resolution (LR) images by sharpening rough object edges and enriching the missing detailed textures \cite{lim2017enhanced,wang2020deepsurvey}. 
SISR has been demonstrated beneficial to other vision tasks, such as classification, detection, and segmentation \cite{dai2016image, shermeyer2019effects, wang2020deep}. 
For the classification tasks, LR images could lose the detailed context information, which leads to significant performance degradation. 
Thus, works have been developed to combine image super-resolution with the typical classification networks to recover the details required for fine-grained image classification \cite{cai2019convolutional,ABDELNASSER201784}. 
For the detection task, to detect small objects from LR images, super-resolution methods have been investigated to improve the performance with SISR as a pre-processing step \cite{shermeyer2019effects} or using feature-level super-resolution to generate features from LR images similar as those of the corresponding HR images \cite{noh2019better,haris2018task}. 
For the segmentation task, in addition to taking the reconstructed HR images as the input \cite{niu2020effective}, the SR network can also be unified in the segmentation network to help extract HR features guided by the fine-grained structural representation from the SISR branch \cite{wang2020dual}. 
Considering the computation efficiency, we follow the spirit of the dual-branch semantic segmentation \cite{wang2020dual} of the small nodules from the CT images by the two-stream mechanism to involve more detailed information. 
However, as the super-resolution learning branch tends to focus on detailed information, the dual-branch network might not be able to extract rich context for the large lesions.

\begin{figure*}[t]
\centering
\includegraphics[width=\textwidth]{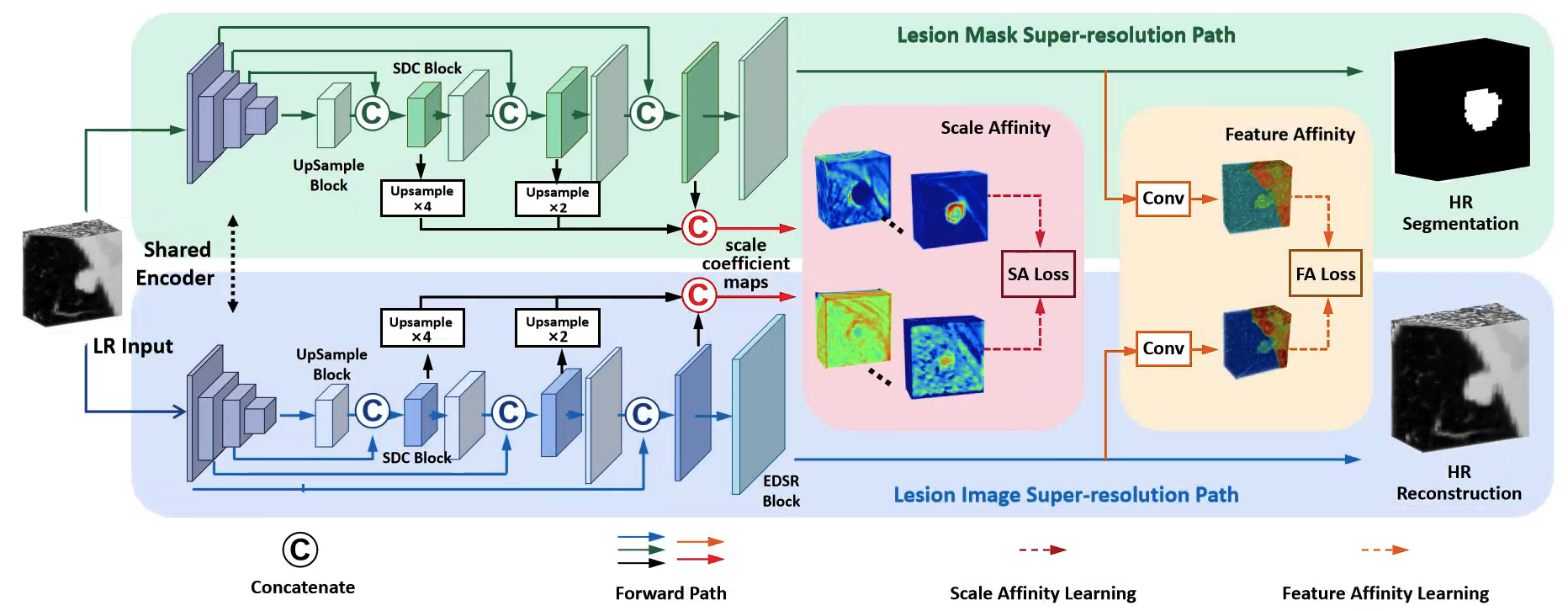}
\caption{Overview of our proposed framework with 3D Unet as the backbone. The lesion mask super-resolution path (LMSR) (\textcolor{green}{in green}) and the lesion image super-resolution path (LISR) (\textcolor{black}{in blue}) share the same encoder and use two decoders of same structure with SDC block. The dual affinity learning modules contain a feature affinity loss computed between features from the two paths (\textcolor{yellow}{in orange}) and a scale affinity loss computed between the scale coefficient maps from the two paths (\textcolor{red}{in red}).}
\label{fig:framework}
\end{figure*}

\section{Methodology}\label{method}
In this section, we first present the design of our proposed scale-aware super-resolution network, including a dual-path super-resolution framework and scale-aware dilated convolution blocks.
Then, we elaborate in order on the feature affinity learning module and the scale affinity learning module.
Finally, we wrap up all proposed components and introduce the optimization procedures for our proposed method.

\subsection{Scale-aware Super-resolution Network}
\subsubsection{Dual-path Super-resolution Framework}
Inspired by \cite{wang2020dual}, we first adopt a multi-task super-resolution (SR) network to improve the tiny lesion segmentation performance with low-resolution medical images.
Specifically, our network consists of two SR paths: a lesion mask super-resolution (LMSR) path for generating high-resolution masks from the low-resolution inputs; and a lesion image super-resolution (LISR) path for reconstructing high-resolution images from the inputs.

As shown in Fig.~\ref{fig:framework}, the two branches share the same encoder and adopt two separated decoders for super-resolution and segmentation. \textcolor{black}{The multi-level features from the encoder are integrated into the decoders by skip connections, following a similar approach to the Unet architecture.
Both paths take as inputs the low-resolution (LR) images and generate high-resolution (HR) outputs of twice the input size.
Different layers are appended to the two decoders to generate the final results, respectively.
Particularly, the HR reconstruction is generated via an enhanced deep super-resolution (EDSR) module \cite{lim2017enhanced} appended to the LISR decoder, and the HR segmentation output is generated by upsampling the feature map from the LMSR decoder twice its original scale.}

Specifically, the LMSR path is trained using the Dice loss:
\begin{equation}\label{dice_loss}
  \mathcal{L}_{\rm LMSR} = 1-\frac{2\sum_{i=1}^{N}p_{i}y_{i}+\xi}{\sum_{i=1}^{N}p_{i}^{2}+\sum_{i}^{N}y_{i}^{2}+\xi }
\end{equation}
where $p_i$ means the $i$-th voxel on the predicted HR segmentation mask, $y_i$ is the $i$-th voxel on the HR segmentation label, $N$ is the total number of voxels, and $\xi$ is a smoothing term to avoid zero division.

In the meantime, the LISR branch is trained with the mean square error (MSE) loss, with the HR images as the groundtruth.
\textcolor{black}{We noticed that the MSE loss could pay too much attention to the large irrelevant background regions, generating smooth and blurred HR reconstructions.
To address this issue, we use the segmentation groundtruth to increase the weight of loss for the target lesion region, which makes the network focus more on reconstructing the foreground information.}
As a result, the loss for the LISR path is as follows:
\begin{equation} \label{mse_loss}
    \mathcal{L}_{\rm LISR} = \frac{1}{N}\left ( \lambda_{1}  \sum_{i \in \Omega}\|q_{i} - x_{i}\|_{2}^{2} + \lambda_{2} \sum_{i \notin \Omega}\|q_{i} - x_{i}\|_{2}^{2} \right )
\end{equation}
where $q_i$ is the voxel on the reconstructed HR images generated by the LISR path, $x_i$ is the voxel on the HR image, $\Omega$ is the voxel set for the lesion region, and $\lambda_{1}$ and $\lambda_{2}$ are two hyperparameters to weigh the losses. We empirically set $\lambda_{1}$ to $0.8$ and $\lambda_{2}$ to $0.2$ throughout our experiments.

With such a dual-path structure, the LISR branch will provide information from higher resolution to the shared encoder, which in turn assists the LMSR branch to extract more detailed information for tiny lesion segmentation.

\subsubsection{Scale-aware Dilated Convolution Block}
The dual-path super-resolution framework largely recovers the information for small lesions.
However, the LISR branch puts more emphasis on the extraction of the detailed low-level features, which could weaken the extraction of the overall semantic information, especially for the large lesions.
Moreover, as lesions are of diverse scales, different receptive fields will also be preferred to extract more precise features.
Therefore, we further introduce scale-aware dilated convolution (SDC) blocks to dynamically adjust the receptive field for lesions of various sizes.

\begin{figure}[t]
  \centering
  \includegraphics[width=0.8\linewidth]{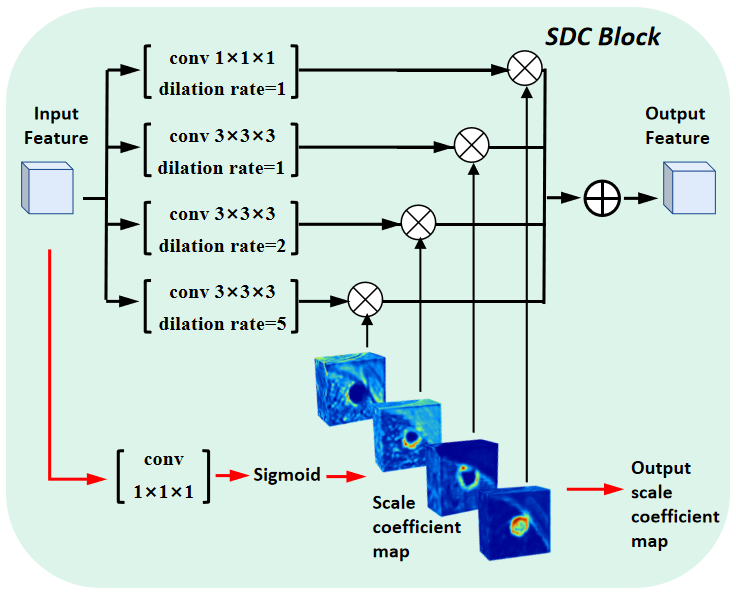}
  \caption{Illustration of a scale-aware dilated convolution (SDC) block. Each SDC block consists of four dilated convolutional kernels with different dilation rates. 
  Then four scale coefficient maps are used to weigh the contributions of the kernels.
  $\otimes$: voxel-wise multiplication; $\oplus$: voxel-wise summation.}
  \label{fig:SDC}
\end{figure}

\textcolor{black}{Specifically, multiple dilated convolutional kernels are stacked in parallel, and the kernels have different dilation rates to generate multi-scale features.
Previous works \cite{chen2017deeplab,zhao2017pyramid} directly fused the multi-scale features with equal importance.
In practice, the features of different objects can be obtained with different receptive fields, and the contributions of the multi-scale features to the final results should not be equal, but rather depend on the target sizes.
In specific, segmenting small (\textcolor{black}{resp.} large) lesions relies more on the features from smaller (\textcolor{black}{resp.} larger) receptive fields.
}
Hence, we measure the contributions of each feature at the voxel level via the proposed scale coefficient maps as follows:
\begin{equation}
  \begin{split}
    &S_{i} = \sigma(Conv_{i}( X_{\rm input})) \\
    X_{\rm output} = &\sum_{i}^{M}(S_{i} \otimes DilatedConv_{i}(X_{\rm input}))
  \end{split}   
\end{equation}
where we call $S_{i}$ a scale coefficient map, $X$ represents the feature maps, $M$ is the number of scale coefficient maps, $\sigma$ represents the sigmoid function, and $\otimes$ means voxel-wise multiplication.
The scale coefficient map softly assigns weights to features abstracted from different dilated convolutions, which allows the model to seek for a more suitable combination of the receptive fields and eventually enables dynamic adjustment according to the inputs.
Fig.~\ref{fig:SDC} illustrates the implementation of a SDC block, where four dilated convolutions and four scale coefficient maps are used to obtain the scale-aware output feature maps.
The SDC blocks are used to replace several convolution blocks of the decoders in both the LMSR and LISR branches, which is also illustrated in Fig.~\ref{fig:framework}.

\subsection{Dual Affinity Learning}
\label{sec:dual affinity}
The dual-path structure and the SDC blocks promote the segmentation for lesions of diverse sizes.
Here, we further enhance the LMSR learning via guidance from the LISR branch by introducing two affinity learning modules.

\subsubsection{Feature Affinity Learning Module}
\textcolor{black}{The LMSR branch generates HR lesion mask and focuses on abstracting semantic information.
Meanwhile, the LISR branch conducts a super resolution task, which learns richer and more detailed HR information than the LMSR branch.}
In light of such an observation~\cite{wang2020dual}, we adopt a feature affinity (FA) module to promote the LMSR branch's segmentation learning with the guidance from the LISR branch.
The FA module aligns the features correspondence between the two branches, facilitating the integration of the detailed HR information into the LMSR branch and benefiting the segmentation task.

\textcolor{black}{Specifically, two convolutional kernels are separately added before the output layers} of the two branches to generate features of dimension $\mathbb{R}^{C\times{W}\times{H}\times{D}}$, with $W$, $H$, and $D$ the spatial dimensions and $C$ the channel dimension.
Next, we conduct adaptive average pooling on the feature maps to downsample them to $\mathbb{R}^{C\times{W'}\times{H'}\times{D'}}$ in order to reduce the computational consumption.
We then reshape the feature map into a 2D tensor $F' \in\mathbb{R}^{C\times(W'H'D')}$ and compute the Gram Matrix $GFA$ for feature affinity learning by:

\begin{equation} \label{gram_matrix_F}
    GFA = {{F}'}^{\rm T} \cdot F'
\end{equation}
%where $F'$ is the pooled feature map. 
where $GFA \in \mathbb{R}^{(W'H'D')\times(W'H'D')}$. 
We empirically set ${W}'=W//16$, ${H}'=H//16$, and ${D}'=D//16$.

The feature affinity loss is optimized by minimizing the similarity between the obtained Gram Matrices from the dual branches, \textcolor{black}{which can be described with the following equation:}
\begin{equation} \label{fa_loss}
    \mathcal{L}_{\rm FA} = \frac{1}{({W}'{H}'{D}')^{2}}\sum_{i=1}^{{W}'{H}'{D}'}\sum_{j=1}^{{W}'{H}'{D}'}\left \| GFA_{ij}^{\rm seg}-GFA_{ij}^{\rm sr} \right \|_{2}^{2} \\
\end{equation}
where $i$ and $j$ are the spatial indices, $GFA^{\rm seg}$ and $GFA^{\rm sr}$ are the Gram Matrices for the features from the LMSR and LISR branches, respectively.

\subsubsection{Scale Affinity Learning Module}
The feature affinity loss imposes such a constraint on the dual-path learning network: the features from the two branches should share the same voxel-wise correlations, despite that the features may obtain different semantics.
However, the optimization objectives are different in the two branches, leading to different emphases on the foregrounds and backgrounds which may cause the instability of the training process. 
We thus migrate this assumption to the scale coefficient maps obtained by our SDC blocks, i.e., we emphasize that the scale coefficient maps from two paths shall also share similar voxel-wise correlations.
To this end, we propose a scale affinity (SA) learning module to enhance the learning between the SDC blocks of the two branches and stabilize the model training.

\textcolor{black}{Particularly, we interpolate all the scale coefficient maps to be the same shape as the largest one and concatenate them to obtain a general map $S_{\rm all}$.}
Similar to the FA module, we then minimize the similarity between the Gram Matrices of scale coefficient maps.
Note that each SDC block contains four scale coefficient maps, and each branch have three SDC blocks, so we have $S_{\rm all}\in\mathbb{R}^{12\times{H}\times{W}\times{D}}$.
We also pool and reshape $S_{\rm all}$ into a 2D tensor of dimension $\mathbb{R}^{12\times({H'}{W'}{D'})}$, denoted as $S'_{\rm all}$, where ${W}'=W//16$, ${H}'=H//16$, and ${D}'=D//16$.
%Suppose $S_{l3}\in\mathbb{R}^{{H}\times{W}}$, then we have $S_{\rmall}\in\mathbb{R}^{12 \times{H}\times{W}}$.
The Gram Matrix of the scale coefficient maps is then computed by the following equation:
\begin{equation} \label{gram_matrix_S}
    GSA = {{S'}_{\rm all}}^{\rm T} \cdot {S'}_{\rm all}
\end{equation}
where $GSA\in\mathbb{R}^{{H'}{W'}{D'} \times {H'}{W'}{D'}}$. The scale affinity loss is then conducted as follows: 
\begin{equation} \label{sa_loss}
    \mathcal{L}_{\rm SA} = \frac{1}{({W'}{H'}{D'})^{2}}\sum_{i=1}^{{W'}{H'}{D'}}\sum_{j=1}^{{W'}{H'}{D'}}\left \| GSA_{ij}^{\rm seg}-GSA_{ij}^{\rm sr} \right \|_{2}^{2}
\end{equation}
where $i$ and $j$ are the spatial indices, $GSA^{\rm seg}$ and $GSA^{\rm sr}$ are the Gram Matrices for the scale coefficient maps from the LMSR and LISR branches, respectively.
\textcolor{black}{Consequently, this affinity loss guides the dual-branch learning with the alignment of receptive fields.}

\subsection{Overall Network and Optimization Procedure}
The complete network is illustrated in Fig.~\ref{fig:framework}, which contains dual paths of super-resolution networks with scale-aware dilated convolution blocks in the decoders. 
Moreover, the dual affinity modules, i.e., the feature affinity module and scale affinity module, are introduced between the two paths to improve the mask super-resolution learning. 
The overall objective of our proposed model is as follows:

\begin{equation}\label{loss_all}
    \mathcal{L} = \mathcal{L}_{\rm LMSR} + \alpha\mathcal{L}_{\rm LISR } + \beta\mathcal{L}_{\rm FA} + \gamma\mathcal{L}_{\rm SA}
\end{equation}
where $\alpha$, $\beta$, and $\gamma$ are the hyper-parameters to weigh the losses.
Throughout our experiments, we empirically set the loss weights to $1$.
The whole objective function can be optimized end-to-end during the training stage.

\section{Experimental Results}\label{experiments}
We conducted experiments on four challenging lesion segmentation datasets, covering two datasets of 3D pulmonary nodule segmentation, one dataset of 2D skin lesion segmentation, and one benchmark of 2D polyp segmentation.
%The number of epochs is 80 in total.
All the networks involved in our experiments were implemented using Pytorch \cite{paszke2019pytorch} on a server with 2 NVIDIA GeForce GTX TITAN XP GPUs (12 GB).

\begin{table*}\caption{Pulmonary nodule segmentation comparison of our proposed method and other state-of-the-art approaches on the MHN and LIDC-IDRI datasets. All models are implemented in 3D. Results are reported for nodules of different diameters and the average over all testing data. The best performance is highlighted in \textbf{bold}.}
\centering
\resizebox{\textwidth}{!}{
\begin{tabular}{c|ccccccccc|ccccccccc}
\hline
\toprule[2pt]
\multirow{3}{*}{} & \multicolumn{9}{c|}{\textbf{MHN}}                                                                                                                                                                                                                                                                                                       & \multicolumn{9}{c}{\textbf{LIDC}}                                                                                                                                                                                                                                                                                                     \\ \cline{2-19} 
                  & \multicolumn{3}{c|}{\textless{} 4 mm}                                                                                        & \multicolumn{3}{c|}{$\geq$ 4 mm}                                                                                        & \multicolumn{3}{c|}{Overall}                                                               & \multicolumn{3}{c|}{\textless{} 4 mm}                                                                                        & \multicolumn{3}{c|}{$\geq$ 4 mm}                                                                                       & \multicolumn{3}{c}{Overall}                                                               \\ \cline{2-19} 
                  & DSC                             & IOU                             & \multicolumn{1}{c|}{MAE}     & DSC                             & IOU                             & \multicolumn{1}{c|}{MAE}     & DSC                             & IOU                             & MAE     & DSC                             & IOU                             & \multicolumn{1}{c|}{MAE}     & DSC                             & IOU                             & \multicolumn{1}{c|}{MAE}    & DSC                             & IOU                             & MAE    \\ \hline
\textbf{Unet \cite{ronneberger2015u}}            & 0.501                           & 0.378                           & \multicolumn{1}{c|}{\textbf{0.001}} & 0.614                           & 0.482                           & \multicolumn{1}{c|}{0.018}   & 0.577                           & 0.448                           & 0.013  & 0.448                           & 0.342                           & \multicolumn{1}{c|}{0.028} & 0.730                           & 0.602                           & \multicolumn{1}{c|}{0.008} & 0.726                           & 0.598                           & 0.008 \\
\textbf{PSPNet \cite{zhao2017pyramid}}           & 0.323                           & 0.223                           & \multicolumn{1}{c|}{0.002}  & 0.569                           & 0.429                           & \multicolumn{1}{c|}{0.016}  & 0.488                           & 0.361                           & \textbf{0.011}  & 0.149                           & 0.088                           & \multicolumn{1}{c|}{0.040}  & 0.653                           & 0.505                           & \multicolumn{1}{c|}{0.009} & 0.644                           & 0.498                           & 0.010 \\
\textbf{Unet++ \cite{zhou2019unet++}}          & 0.497                           & 0.373                           & \multicolumn{1}{c|}{\textbf{0.001}} & 0.639                           & 0.506                           & \multicolumn{1}{c|}{0.015}   & 0.592                           & 0.462                           & \textbf{0.011}   & 0.440                           & 0.328                           & \multicolumn{1}{c|}{0.029} & 0.738                           & 0.609                           & \multicolumn{1}{c|}{\textbf{0.007}} & 0.733                           & 0.604                           & 0.008 \\
\textbf{Unet3+  \cite{huang2020unet}}            & 0.530                           & 0.394                           & \multicolumn{1}{c|}{\textbf{0.001}}   & 0.625                           & 0.487                           & \multicolumn{1}{c|}{0.018}   & 0.597                           & 0.459                           & 0.012   & 0.412                           & 0.307                           & \multicolumn{1}{c|}{0.032} & 0.746                           & 0.620                           & \multicolumn{1}{c|}{\textbf{0.007}} & 0.741                           & 0.615                           & 0.008 \\
\textbf{nnUnet  \cite{isensee2021nnu}}           & \textbf{0.556}                           & \textbf{0.416}                           & \multicolumn{1}{c|}{\textbf{0.001}}  & 0.628                           & 0.489                           & \multicolumn{1}{c|}{0.019}  & 0.605                           & 0.465                           & 0.013  & 0.529                           & 0.424                           & \multicolumn{1}{c|}{\textbf{0.020}}  & 0.728                           & 0.599                           & \multicolumn{1}{c|}{\textbf{0.007}}  & 0.725                           & 0.596                           & 0.008 \\
\textbf{UNETR  \cite{hatamizadeh2022unetr}}            & 0.408                           & 0.291                           & \multicolumn{1}{c|}{\textbf{0.001}}  & 0.428                           & 0.306                           & \multicolumn{1}{c|}{0.024}  & 0.421                           & 0.301                           & 0.016  & 0.208                           & 0.147                           & \multicolumn{1}{c|}{0.035}  & 0.707                           & 0.576                           & \multicolumn{1}{c|}{0.008}  & 0.700                           & 0.569                           & 0.009  \\
\textbf{DSRL \cite{wang2020dual}}              & 0.546 & 0.406 & \multicolumn{1}{c|}{\textbf{0.001}} & 0.587                           & 0.447                           & \multicolumn{1}{c|}{0.020} & 0.568                          & 0.428                          & 0.014 & 0.291                           & 0.211                           & \multicolumn{1}{c|}{0.030} & 0.616                           & 0.482                           & \multicolumn{1}{c|}{0.013} & 0.611                           & 0.478                          & 0.013 \\ 
\textbf{\textcolor{black}{PFSeg \cite{wang2021patch}}}             & \textcolor{black}{0.530} & \textcolor{black}{0.390} & \multicolumn{1}{c|}{\textbf{\textcolor{black}{0.001}}} & \textcolor{black}{0.605}                           & \textcolor{black}{0.467}                           & \multicolumn{1}{c|}{\textbf{\textcolor{black}{0.013}}} & \textcolor{black}{0.586}                           & \textcolor{black}{0.447}                           & \textcolor{black}{0.013} & \textcolor{black}{0.496}                           & \textcolor{black}{0.403}                           & \multicolumn{1}{c|}{\textcolor{black}{0.022}} & \textcolor{black}{0.752}                           & \textcolor{black}{0.624}                           & \multicolumn{1}{c|}{\textbf{\textcolor{black}{0.007}}} & \textcolor{black}{0.748}                           & \textcolor{black}{0.620}                           & \textcolor{black}{0.008} \\ \hline
\textbf{Our Method}        & 0.536                           & 0.395                           & \multicolumn{1}{c|}{0.002}  & \textbf{0.648} & \textbf{0.507} & \multicolumn{1}{c|}{\textbf{0.013}}   & \textbf{0.611} & \textbf{0.470} & \textbf{0.011}  & \textbf{0.560} & \textbf{0.449} & \multicolumn{1}{c|}{0.022}  & \textbf{0.760} & \textbf{0.630} & \multicolumn{1}{c|}{\textbf{0.007}} & \textbf{0.757} & \textbf{0.627} & \textbf{0.007} \\ 
\bottomrule[2pt]
\end{tabular}
}
\label{compare with SOTA in NODULE}
\end{table*}

\begin{figure*}[t]
  \centering
  \includegraphics[width=\linewidth]{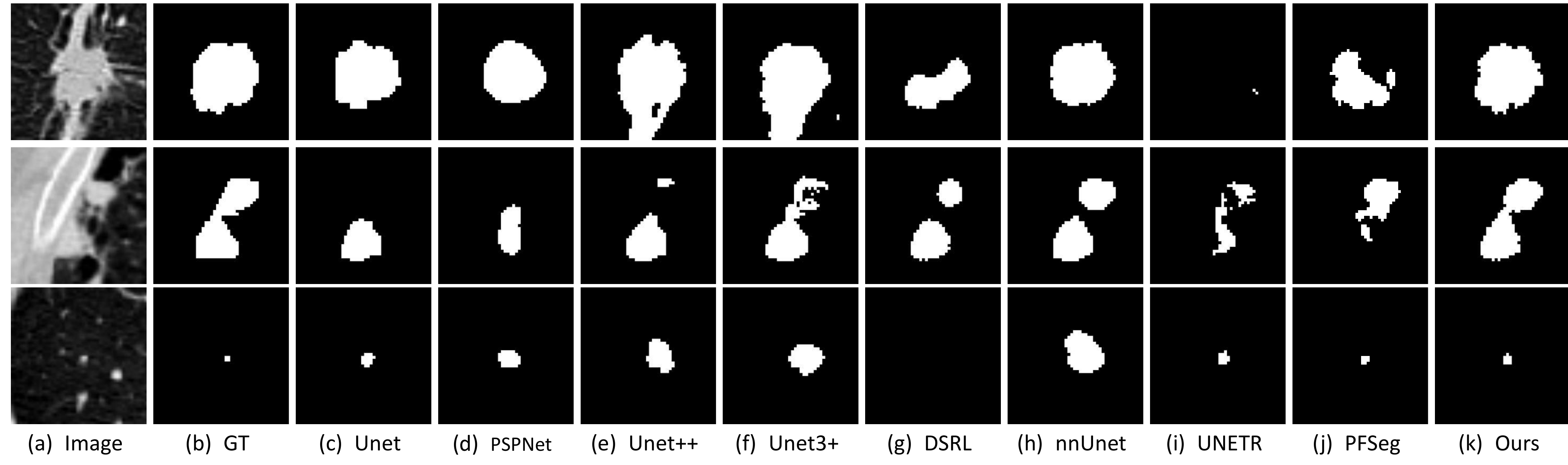}
  \caption{Comparison of the qualitative nodule segmentation results among other SOTA approaches and our method. Three nodules of different sizes are displayed at their largest slices and the ground-truth, as well as the segmented masks of the compared methods, are shown.}
  \label{fig:seg_nodule}
\end{figure*}

\subsection{Lung Nodule Segmentation}

\subsubsection{Datasets} 
Two datasets for nodule segmentation were evaluated: the MHN dataset and the LIDC-IDRI \cite{armato2011lung} dataset.
The MHN dataset is a self-collected multi-hospital dataset including 2,047 CT scans with 17,769 nodules annotated \textcolor{black}{with} masks. 
\textcolor{black}{The nodules were localized by the agreement of at least two radiologists from a cohort of four radiologists.
After the locations of nodules were determined, the pixel-level binary masks were then delineated by one radiologist.}
The dataset was then split into training set (1,532 CT scans with 13,905 nodules), validation set (250 CT scans with 2,068 nodules), and test set (265 CT scans with 1,796 nodules).

The LIDC-IDRI dataset labeled nodules of diameter above 3 mm with volumetric contours.
The CT scans are collected from seven academic centers and eight medical imaging companies and the masks are annotated by four experienced radiologists. 
We kept the samples with mask annotations while excluding several samples with excessive slice thickness or poor scan reconstruction.
As a result, we obtained 872 CT scans with 2,613 nodules and split the data from the nodule level to training (1,888 nodules), validation (211 nodules), and testing (514 nodules) sets. 

\subsubsection{Implementation Details}
For both datasets, we cropped patches around the nodule regions for better comparison of different methods and all chest CT spacing to 1/1.2 mm.
Specifically, we cropped patches of size $96\times 96\times 96$ as the LISR groundtruth and used the corresponding mask patches as the LMSR groundtruth.
We downsampled the CT patches into the size of $48\times 48\times 48$ as the training inputs.
During testing, we fed the network with $48\times 48\times 48$ patches \textcolor{black}{directly cropped from the original CT images} without downsampling.
We downsample the output predicted mask to keep the size of the output the same as the input patches.
For other compared models, we cropped $48\times 48\times 48$ patches from the CT images as both training and testing data.
Hence, all the models are evaluated with the input data and masks in the same resolution.

\textcolor{black}{According to the Lung Imaging Reporting and Data System (Lung-RADS) by The American College of Radiology, newly developed solid nodules would be assessed as Lung-RADS 2 if less than 4 mm and higher Lung-RADS scores otherwise \cite{American2019Lung}. Therefore, we used 4 mm as a boundary value to evaluate the performance of the methods on the lesions of different scales.}

\begin{table*}[]\caption{Skin lesion segmentation comparison of our proposed method and other state-of-the-art approaches on the ISIC2018 dataset. Results are reported in Dice Similarity Coefficients for lesion of different pixels and the average over all testing data. The best performance is highlighted in \textbf{bold}.}
\centering
\begin{tabular}{c|cccccccccccc}
\hline
\toprule[2pt]
                               & \multicolumn{9}{c}{\textbf{ISIC2018}}                                                                                                                                                                                                                                                                                                   \\ \cline{2-10} 
                               & \multicolumn{3}{c|}{\textless 100 pixels}                                                                                & \multicolumn{3}{c|}{$\geq$ 100 pixels}                                                                  & \multicolumn{3}{c}{Overall}                                                                \\ \cline{2-10} 
\multirow{-3}{*}{}             & DSC                          & IOU                          & \multicolumn{1}{c|}{MAE}                         & DSC                          & IOU                          & \multicolumn{1}{c|}{MAE}                         & DSC                          & IOU                          & MAE                         \\ \hline
\textbf{Unet \cite{ronneberger2015u}}                          & 0.820                        & 0.743                        & \multicolumn{1}{c|}{0.022}                        & 0.848                        & 0.767                        & \multicolumn{1}{c|}{0.077}                        & 0.843                        & 0.762                        & 0.066                        \\
\textbf{PSPNet \cite{zhao2017pyramid}}                        & 0.809                        & 0.731                        & \multicolumn{1}{c|}{0.063}                        & 0.898                        & 0.828                        & \multicolumn{1}{c|}{0.066}                        & 0.880                        & 0.809                        & 0.065                        \\
\textbf{Unet++ \cite{zhou2019unet++}}                         & 0.848                        & 0.762                       & \multicolumn{1}{c|}{0.022}                        & 0.883                        & 0.804                        & \multicolumn{1}{c|}{0.064}                        & 0.876                        & 0.796                        & 0.056                        \\
\textbf{Unet3+ \cite{huang2020unet}}                         & 0.882                        & 0.812                        & \multicolumn{1}{c|}{0.014}                        & 0.875                        & 0.805                        & \multicolumn{1}{c|}{0.063}                        & 0.876                        & 0.806                        & 0.053                        \\
\textbf{nnUnet \cite{isensee2021nnu}}                         & 0.854                        & 0.781                        & \multicolumn{1}{c|}{0.028}                        & 0.888                        & 0.819                        & \multicolumn{1}{c|}{0.058}                         & 0.877                        & 0.806                        & 0.049                         \\
\textbf{TransUnet \cite{chen2021transunet}}                      & 0.830                        & 0.735                        & \multicolumn{1}{c|}{0.025}                        & 0.905                        & 0.835                        & \multicolumn{1}{c|}{0.052}                         & 0.890                        & 0.817                        & 0.047                         \\
\textbf{DSRL \cite{wang2020dual}}                           & 0.889                        & 0.813                        & \multicolumn{1}{c|}{0.043}                        & 0.822                        & 0.721                        & \multicolumn{1}{c|}{0.087}                        & 0.833                        & 0.735                        & 0.078                        \\ 
\textbf{\textcolor{black}{PFSeg \cite{wang2021patch}}}                           & \textcolor{black}{0.861}                        & \textcolor{black}{0.784}                        & \multicolumn{1}{c|}{\textcolor{black}{0.025}}                        & \textcolor{black}{0.901}                        & \textcolor{black}{0.828}                        & \multicolumn{1}{c|}{\textcolor{black}{0.057}}                        & \textcolor{black}{0.893}                        & \textcolor{black}{0.820}                        & \textcolor{black}{0.051}                        \\ \hline 
\textbf{Our Method}                     & \textbf{0.902}                        & \textbf{0.829}                        & \multicolumn{1}{c|}{\textbf{0.013}}                        & \textbf{0.913}                        & \textbf{0.848}                        & \multicolumn{1}{c|}{\textbf{0.049}}                        & \textbf{0.911}                        & \textbf{0.844}                        & \textbf{0.042}                        \\ \hline
\bottomrule[2pt]
\end{tabular}
\label{compare with SOTA skin lesion}
\end{table*}

\begin{figure*}[t]
  \centering
  \includegraphics[width=\linewidth]{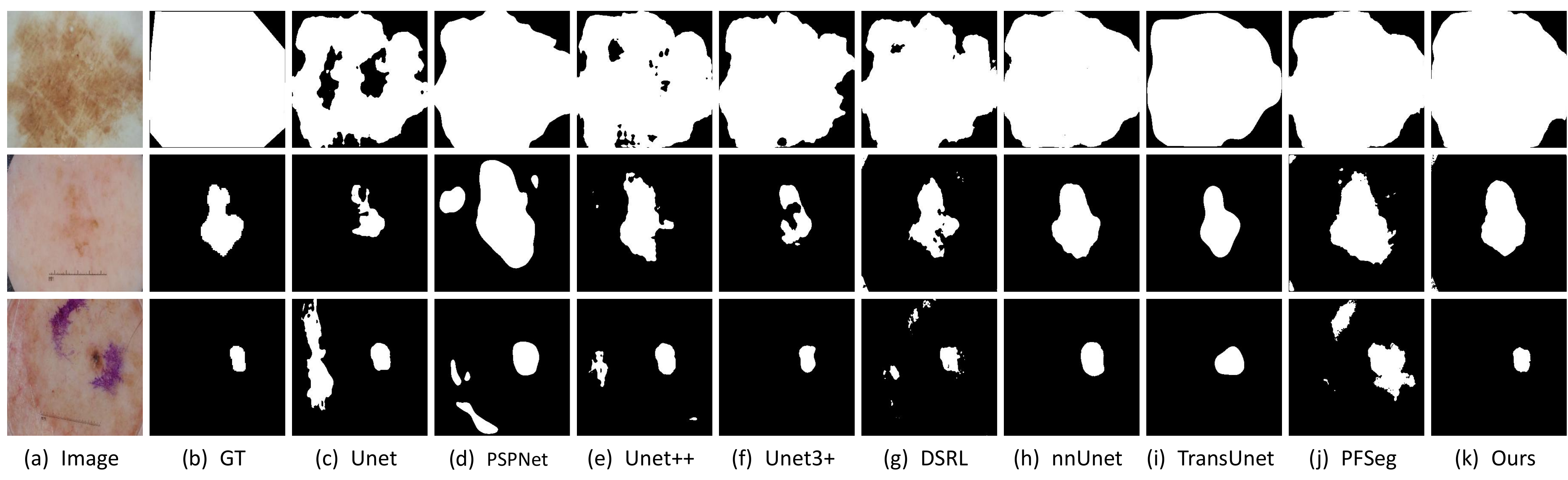}
  \caption{Qualitative skin lesion segmentation results for ISIC2018 comparison of other SOTA approaches and our method. Three samples of the skin lesions of different sizes are displayed and the ground-truth, as well as the segmented masks of compared methods, are shown.}
  \label{fig:seg_isic}
\end{figure*}

\subsubsection{Comparison with the State-of-the-arts}
We compared our method against the 3D implementations of other  state-of-the-art segmentation models, including PSPNet\cite{zhao2017pyramid}, Unet++\cite{zhou2019unet++}, Unet3+ \cite{huang2020unet}, nnUnet \cite{isensee2021nnu}, UNETR \cite{hatamizadeh2022unetr}, \textcolor{black}{as well as two multi-task networks, DSRL \cite{wang2020dual} and PFSeg \cite{wang2021patch}}.
We applied three metrics, including Dice similarity coefficient (DSC), Intersection over Union (IOU), and mean absolute error (MAE), to evaluate the segmentation performance.

% Two datasets for nodule segmentation , 
Both the MHN and the LIDC-IDRI datasets were included in the experiments.
Each dataset was split into two groups according to the lesion sizes, and the results are shown in Table \ref{compare with SOTA skin lesion}. 
On the MHN dataset, although the performance on the tiny nodules has a certain trade-off, our model showed significant overall improvement with at least 0.6\% in DSC, 0.5\% in IOU, and achieves the best MAE compared to other methods. 
On the LIDC-IDRI dataset, our proposed method achieved an average DSC of 75.7\% in the testing set, which was 3.1\% higher in DSC, 2.9\% higher in IOU than Unet, and 0.9\% higher in DSC and 0.7\% in IOU than DSRL.
For nodules smaller than 4 mm, our method outperformed other methods with at least 3.1\% in DSC and 2.5\% in IOU.
For nodules larger than 4 mm, our method improved at least 0.8\% in DSC and 0.6\% in IOU.

\textcolor{black}{Our method did not achieve the best performance on tiny nodules ($<$4 mm) from the MHN dataset, compared with nnUnet. One possible reason is the that we downsampled the original CT crops as the inputs for our method. As the tiny nodules smaller than 4 mm may only exhibit a few pixels on the original CT scan, our downsampling strategy could lose some original information despite the super-resolution training. Nevertheless, our main goal is to obtain SOTA overall lesion segmentation performance. Our model showed a better trade-off on the performance on the small or large lesions and achieved the best overall performance, demonstrating its effectiveness on segmenting lesions across different scales.}

\textcolor{black}{We also show in Fig. \ref{fig:seg_nodule} the qualitative results of ours as well as eight other approaches.}
As can be observed, our method consistently attained better visual segmentation results for nodules of all different scales.
Moreover, for hard cases, such as the nodules with irregular shapes (the 2nd row) or extremely tiny nodules (the 3rd row), our method also demonstrates more robust results.

\subsection{Skin Lesion Segmentation}
\subsubsection{Dataset and Implementation Details}
%\lyluo{number} 
The ISIC2018 skin lesion segmentation dataset \cite{codella2019skin} contains 2,694 pictures of various-size skin lesions. We used 1,617 images for training, 539 for validation, and 538 for testing. 
We resized the original images to $640\times 640$ to be the HR ground-truth for the LISR path.
The annotation masks were resized to $640\times640$ as the ground-truth for semantic segmentation.
We also resized the images to be $320 \times 320$ as the LR inputs.
The LR images are also used as the inputs of the other networks.
As the physical measurements of the lesions were not recorded, we measured the lesion scales in pixels on the $320\times 320$ images.
\textcolor{black}{We used a boundary value of 100 pixels to separate the data into two groups for evaluating the performance on different lesion scales.}

\begin{table*}[]\caption{Polyp segmentation comparison of our proposed method and other state-of-the-art approaches on a multi-source dataset, containing data from Kvasir, CVC-612, ColonDB, ETIS, and CVC-T. PraNet* means the PraNet with a ResNet-101 backbone. The best performance and the second best performance are highlighted in \textbf{bold} and \underline{underlined}, respectively.}
\centering
\begin{tabular}{c|cllcllcllcllcll}
\hline
\toprule[2pt]
\multirow{3}{*}{\textbf{Method}} & \multicolumn{15}{c}{\textbf{Polyp segmentation}}                                                                                                                                                                                                                                                                                                 \\ \cline{2-16} 
                                 & \multicolumn{3}{c|}{\textbf{Kvasir}}                                  & \multicolumn{3}{c|}{\textbf{CVC-612}}                                 & \multicolumn{3}{c|}{\textbf{ColonDB}}                                 & \multicolumn{3}{c|}{\textbf{ETIS}}                                    & \multicolumn{3}{c}{\textbf{CVC-T}}               \\ \cline{2-16} 
                                 & \multicolumn{1}{l}{DSC} & IOU            & \multicolumn{1}{l|}{MAE}   & \multicolumn{1}{l}{DSC} & IOU            & \multicolumn{1}{l|}{MAE}   & \multicolumn{1}{l}{DSC} & IOU            & \multicolumn{1}{l|}{MAE}   & \multicolumn{1}{l}{DSC} & IOU            & \multicolumn{1}{l|}{MAE}   & \multicolumn{1}{l}{DSC} & IOU            & MAE   \\ \hline
\textbf{Unet \cite{ronneberger2015u}}                   & 0.818                   & 0.746          & \multicolumn{1}{l|}{0.055} & 0.823                   & 0.755          & \multicolumn{1}{l|}{0.019} & 0.512                   & 0.444          & \multicolumn{1}{l|}{0.061} & 0.398                   & 0.335          & \multicolumn{1}{l|}{0.036} & 0.710                   & 0.627          & 0.022 \\
\textbf{Unet++ \cite{zhou2019unet++}}                 & 0.821                   & 0.743          & \multicolumn{1}{l|}{0.048} & 0.794                   & 0.729          & \multicolumn{1}{l|}{0.022} & 0.483                   & 0.410          & \multicolumn{1}{l|}{0.064} & 0.401                   & 0.344          & \multicolumn{1}{l|}{0.035} & 0.707                   & 0.624          & 0.018 \\
\textbf{nnUnet \cite{isensee2021nnu}}                  & 0.876                   & 0.816          & \multicolumn{1}{l|}{\textbf{0.015}} & 0.859                   & 0.804          & \multicolumn{1}{l|}{0.015} & 0.733                   & 0.657 & \multicolumn{1}{l|}{0.047} & 0.633                   & 0.568 & \multicolumn{1}{l|}{0.029} & 0.843                   & 0.763          & 0.012 \\
\textbf{TransUnet \cite{chen2021transunet}}                & 0.751                   & 0.647          & \multicolumn{1}{l|}{0.077} & 0.582                   & 0.484          & \multicolumn{1}{l|}{0.086} & 0.539                   & 0.436          & \multicolumn{1}{l|}{0.081} & 0.292                   & 0.234          & \multicolumn{1}{l|}{0.091} & 0.714                   & 0.601          & 0.036 \\

\textbf{\textcolor{black}{PFSeg \cite{wang2021patch}}}                & \textcolor{black}{0.787}                  & \textcolor{black}{0.705}          & \multicolumn{1}{l|}{\textcolor{black}{0.044}} & \textcolor{black}{0.673}                   & \textcolor{black}{0.572}          & \multicolumn{1}{l|}{\textcolor{black}{0.053}} & \textcolor{black}{0.656}                   & \textcolor{black}{0.562}          & \multicolumn{1}{l|}{\textcolor{black}{0.059}} & \textcolor{black}{0.607}                   & \textcolor{black}{0.517}          & \multicolumn{1}{l|}{\textcolor{black}{0.058}} & \textcolor{black}{0.615}                   & \textcolor{black}{0.523}          & \textcolor{black}{0.056} \\

\textbf{PraNet \cite{fan2020pranet}}                  & 0.898              
& 0.840 & \multicolumn{1}{l|}{0.030} & 0.899                   & 0.849 & \multicolumn{1}{l|}{\underline{0.009}} & 0.709                   & 0.640          & \multicolumn{1}{l|}{0.043} & 0.628                   & 0.567          & \multicolumn{1}{l|}{0.031} & 0.871                   & 0.797          & \underline{0.010} \\ 

\textbf{\textcolor{black}{PraNet* \cite{fan2020pranet}}}                  & \textcolor{black}{0.896}  & \textcolor{black}{0.836} & \multicolumn{1}{l|}{\textcolor{black}{0.034}} & \textcolor{black}{0.907}                   & \textcolor{black}{0.850} & \multicolumn{1}{l|}{\textcolor{black}{0.011}} & \textcolor{black}{0.748}                   & \textcolor{black}{0.663}          & \multicolumn{1}{l|}{\underline{\textcolor{black}{0.040}}} & \textcolor{black}{0.709}                   & \textcolor{black}{0.621}          & \multicolumn{1}{l|}{\textcolor{black}{0.039}} & \textcolor{black}{0.841 }                 & \textcolor{black}{0.759 }         & \textcolor{black}{0.014} \\ 

\textbf{\textcolor{black}{MSNet \cite{zhao2021automatic}}}                  & \textbf{\textcolor{black}{0.907}}              
& \textbf{\textcolor{black}{0.862}} & \multicolumn{1}{l|}{\textcolor{black}{0.028}} & \textbf{\textcolor{black}{0.921}}                   & \textbf{\textcolor{black}{0.879}} & \multicolumn{1}{l|}{\textbf{\textcolor{black}{0.008}}} & \underline{\textcolor{black}{0.755}}                   & \underline{\textcolor{black}{0.678}}          & \multicolumn{1}{l|}{\textcolor{black}{0.041}} & \underline{\textcolor{black}{0.710}}                   & \textbf{\textcolor{black}{0.664}}          & \multicolumn{1}{l|}{\underline{\textcolor{black}{0.020}}} & \textcolor{black}{0.869}                   & \textcolor{black}{0.807}          & \underline{\textcolor{black}{0.010}} \\ \hline

\textbf{PraNet+Ours}             & \underline{0.903}          & \underline{0.849}          & \multicolumn{1}{l|}{\underline{0.027}} & 0.899          & 0.849          & \multicolumn{1}{l|}{0.011} & 0.748          & 0.667          & \multicolumn{1}{l|}{0.045} & 0.659          & 0.584          & \multicolumn{1}{l|}{0.056} & \underline{0.874}          & \underline{0.809} & 0.014 \\

\textbf{\textcolor{black}{PraNet*+Ours}}             & \underline{\textcolor{black}{0.903}}          & \textcolor{black}{0.846}          & \multicolumn{1}{l|}{\underline{\textcolor{black}{0.027}}} & \underline{\textcolor{black}{0.913}}          & \underline{\textcolor{black}{0.864}}          & \multicolumn{1}{l|}{\textbf{\textcolor{black}{0.008}}} & \textbf{\textcolor{black}{0.762}}          & \textbf{\textcolor{black}{0.686}}          & \multicolumn{1}{l|}{\textbf{\textcolor{black}{0.035}}} & \textbf{\textcolor{black}{0.726}}          & \underline{\textcolor{black}{0.655}}          & \multicolumn{1}{l|}{\textbf{\textcolor{black}{0.012}}} & \textbf{\textcolor{black}{0.895}}          & \textbf{\textcolor{black}{0.820}} & \textbf{\textcolor{black}{0.007}} \\

 \hline
\bottomrule[2pt]
\end{tabular}
\label{Polyp result}
\end{table*}

\begin{figure*}[t]
  \centering
  \includegraphics[width=\linewidth]{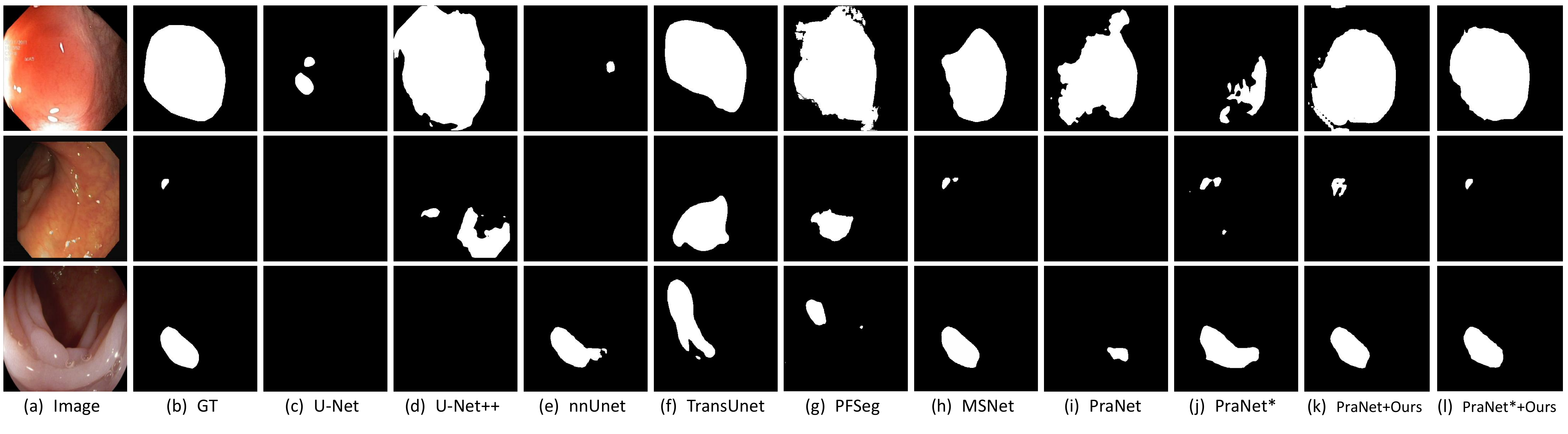}
  \caption{Qualitative Polyp lesion segmentation comparison of other approaches and our method. \textcolor{black}{Three samples of large, middle-size and small lesions
  % Five samples from the five datasets respectively 
  are displayed and the ground truth}, as well as the segmented masks of compared methods, are shown.}
  \label{fig:seg_polyp}
% \end{figure}
\end{figure*}

\subsubsection{Experimental Results}
In Table \ref{compare with SOTA skin lesion}, we report the performance on the two groups according to lesion size and the overall performance with DSC, IOU, and MAE.
We compared our results with eight SOTA segmentation models: Unet, PSPNet, Unet++, Unet3+, nnUnet, TransUnet \cite{chen2021transunet}, DSRL, \textcolor{black}{and PFSeg}.
Our method outperformed other methods by at least 1.3\% in DSC, 1.6\% in IOU with the best MAE for lesions smaller than 100 pixels, and at least 0.8\% in DSC, 1.3\% in IOU and 0.3\% in MAE for lesions larger than 100 pixels. 
The promotion showed the effectiveness of our method on both the small and large lesions. 
\textcolor{black}{Evaluated on all lesions, our method achieved the best overall DSC of 91.1\% with at least 1.8\% improvement over others, the best IOU of 84.4\% with at least 2.4\% improvement, and the best MAE of 4.2\%.}

Qualitative results are illustrated in Fig. \ref{fig:seg_isic}. 
It can be seen that our proposed method delineated the ambiguous lesion edges more accurately and filled the cavity in the middle of the large lesions. 
Meanwhile, our method could reduce many false-positive predictions for large lesions despite the interference of the background objects as shown in the 3rd row.

\subsection{Polyp Segmentation}
\subsubsection{Datasets and Implementation Details}
This experiment was conducted on the well-established benchmark of polyp segmentation \cite{fan2020pranet} with five polyp segmentation datasets: CVC-ClinicDB / CVC-612 \cite{bernal2015wm}, Kvasir \cite{pogorelov2017kvasir}, ETIS \cite{silva2014toward}, CVC-ColonDB \cite{tajbakhsh2015automated}, and EndoScene \cite{vazquez2017benchmark}. 
Following \cite{fan2020pranet}, 90\% of the images from the Kvasir and CVC-612 datasets were used as the training set (1,450 images) and the remained data formed the testing sets (100 images from Kavisir, 62 images from CVC-612, 196 images from ETIS, 380 images from CVC-ColonDB, and 60 images from EndoScene).
\textcolor{black}{To be consistent with previous studies, we focused on comparing the overall segmentation results.}
We used PraNet \cite{fan2020pranet} as a baseline as well as the backbone of our proposed framework.
The images from all the datasets were resized to 704 $\times$ 704 as the HR images and 352 $\times$ 352 as the LR images. 
Three metrics were adopted as that in the PraNet: DSC, mean IOU, and MAE.

\subsubsection{Experimental Results}
\textcolor{black}{Table \ref{Polyp result} compares our method with other competitive methods on the polyp segmentation leaderboard, e.g., PraNet \cite{fan2020pranet} and MSNet \cite{zhao2021automatic}, as well as several SOTA medical image segmentation models, such as Unet, Unet++, nnUnet, TransUnet, and PFSeg.
We also changed the backbone of PraNet from Res2Net-50 to Res2Net-101 \cite{gao2019res2net} and denoted it as PraNet*.
Our method achieved competitive results on the benchmark, showing the best DCS performance on three datasets, the best IOU on two datasets, and the best MAE on four datasets.}
These results demonstrated that our proposed components could be flexibly migrated to other state-of-the-art methods and obtain consistent performance promotion.

The qualitative results are illustrated in Fig \ref{fig:seg_polyp}. Different cases with large ranges of sizes were segmented correctly with our method.
For the tiny lesions shown in row 2, the other methods could not segment the target due to the easily ignorable information as well as the similarity of the foreground and background.
Instead, our method could correctly perceive and segment the target.
The large lesion in row 1 without a clear boundary was also delineated correctly by our method, while other methods may generate incomplete or excessive predictions of the polyp.

\subsection{Analysis of the Method}

\begin{table*}[h]\caption{Ablation study of different component on the MHN dataset. Results are reported in Dice Similarity Coefficients. The best performance is highlighted in \textbf{bold}.}
\centering
\resizebox{0.8\linewidth}{!}{
\begin{tabular}{cccc|ccc|ccc}
\hline
\toprule[2pt]
\multicolumn{4}{c|}{} & \multicolumn{3}{c|}{Lung Nodule Segmentation (MHN)}       & \multicolumn{3}{c}{Skin Lesion Segmentation (ISIC)}                      \\ \hline
SDC  & DSR  & FA & SA & \textless{} 4 mm & $\geq${} 4 mm & Overall        & \textless{} 100 pixels & $\geq$ 100 pixels & Overall        \\ \hline
     &      &    &    & 0.501          & 0.614             & 0.577          & 0.820                 & 0.848                    & 0.843          \\
\checkmark    &      &    &    & 0.479          & 0.637             & 0.586          & 0.795                 & 0.872                    & 0.856          \\
     & \checkmark    &    &    & 0.532          & 0.610             & 0.584          & 0.869                 & 0.801                    & 0.812          \\
\checkmark    & \checkmark    &    &    & 0.532          & 0.621             & 0.590          & 0.836                 & 0.886                    & 0.876          \\
\checkmark    & \checkmark    & \checkmark  &    & \textbf{0.538} & 0.621             & 0.593          & 0.875                 & 0.897                    & 0.893          \\
\checkmark    & \checkmark    &    & \checkmark  & 0.534          & 0.622             & 0.595          & 0.859                 & 0.892                    & 0.886          \\
\checkmark    & \checkmark    & \checkmark  & \checkmark  & 0.536          & \textbf{0.648}    & \textbf{0.611} & \textbf{0.902}        & \textbf{0.913}           & \textbf{0.911} \\ 
\bottomrule[2pt]
\end{tabular}
}
\label{ablation study}
\end{table*}

\subsubsection{Ablation Study on the Proposed Components}

As shown in Table \ref{ablation study}, \textcolor{black}{the SDC blocks enabled adaptive adjustment of the receptive fields, which leads to an improvement of 0.9\% in the average DSC and 2.3\% promotion for nodules larger than 4 mm. }
With the dual-path super-resolution (DSR) structure, the performance for the small nodules improved largely by 3.1\% in DSC, while that of large nodules showed relative drops.
Combining both SDC and DSR further boosted the performance with a better trade-off on nodules of different sizes.
Moreover, incorporating either the FA module or SA module alone brought similar improvement as they both introduced a matching mechanism, improving robust learning by information flow between the dual branches and better balancing the performance among different groups.
Further, the SA module could generate better-aligned features by matching the receptive fields, which stabilizes the mutual learning of the two branches.
Therefore, our full method finally achieved the best overall performance with 3.4\% improvement in DSC compared to the baseline.

We also conducted the ablation study on the ISIC dataset. 
The improvement of 2.4\% for lesions with diameters longer than 100 pixels showed the ability to adjust the receptive fields of the SDC block. 
Meanwhile, the improvement of 4.9\% for lesions with diameters shorter than 100 pixels illustrated the effectiveness of the DSR module, while the performance of larger lesions dropped relatively in consistent with that observed in the MHN dataset.
With both SDC and DSR, the model could balance the performance of the lesions with different sizes and achieve a higher overall DSC score of 87.6\%.
Meanwhile, the FA module and SA module alone could also bring promotion on overall DSC of 1.7\% and 1.0\%, respectively, by enabling information flow between the two branches.
\textcolor{black}{The combination of the two affinity learning modules brought 3.5\% promotion.}
Combining all components, our full model showed the best segmentation performance for skin lesions of various sizes.

\textcolor{black}{These results showed that our proposed method could achieve better adaptability for various sizes of target objects, with effective improvement in the segmentation performance from small lesions to large lesions.}

\begin{figure}[t]
  \centering
  \includegraphics[width=\linewidth]{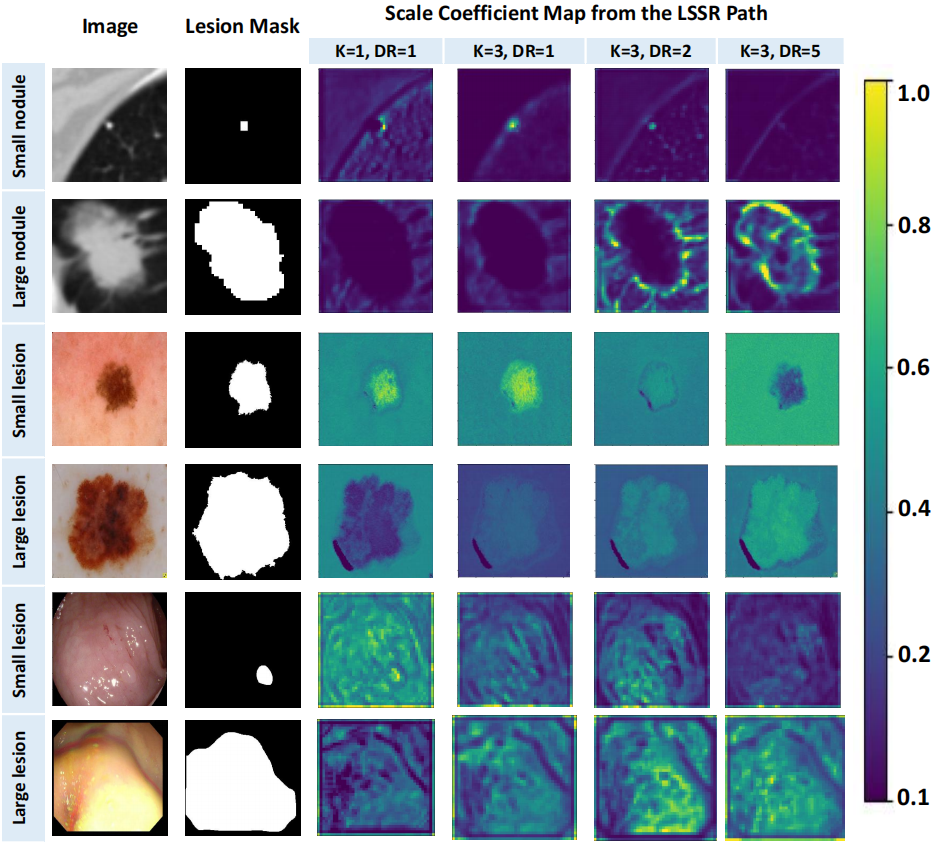}
  \caption{The scale coefficient maps from the highest-level SDC block in decoder of semantic segmentation path for the nodule segmentation task. K: kernel size; DR: dilation rate.}
  \label{fig:score_map}
\end{figure}

\subsubsection{Analysis of Scale Coefficient Map}
We further provide a qualitative analysis by visualizing the scale coefficient maps from the LMSR path.
Specifically, we extracted the maps of the four dilated convolutions from the last SDC block.
Fig. \ref{fig:score_map} shows the scale coefficient maps for lesions of different sizes from the nodule segmentation dataset, the skin lesion dataset, and the polyp segmentation dataset. 
\textcolor{black}{As can be observed, for the small lesions, the scale coefficient maps correctly assigned more weights for the features generated with smaller receptive fields.
For larger lesions, the maps adjusted the weights assigning strategy, so that the features from larger receptive fields contributed more to the final results.}
Moreover, all the maps correctly attended to the foreground objects with less focus on the backgrounds.
These results demonstrated the effectiveness of the proposed scale coefficient maps in dynamically adjusting the receptive field, which also explained the success of our methods in the previous experiments.

\section{Discussion}\label{discussion}
Lesion segmentation is a significant task in computer-assisted diagnosis for multiple diseases as it can provide quantitative estimation for the doctors in primary analysis, treatment, and follow-up process.
However, suffering from the great variation in size of lesions and the low resolution of many medical images, the task remains a challenge in practice. 

Drawing spirit from the utilization of super resolution by other vision tasks, we extended the traditional segmentation structure with an auxiliary branch for image super resolution to improve the performance of the diversely sized lesions segmented from the low resolution medical images.
To further ensure the segmentation result for large lesions in the same time, the SDC blocks were utilized instead of the traditional convolution blocks.
Moreover, we proposed a dual affinity learning scheme including FA and SA to guide the flow of high-resolution information from the LISR branch to the LMSR branch.
The FA module minimized the loss between the Gram matrices of features from both branches, and the SA module minimized the loss between the Gram matrices of scale coefficient maps from both branches. 
Considering the different purpose of the two branches, we utilized the Gram matrices in affinity learning instead of directly optimizing the consistency of the features or scale coefficient maps.

In the FA module, the Gram matrix revealed the relationship between the features, which helped to extract the intrinsic relation and semantic information from LISR and LMSR branches.
\textcolor{black}{However, FA alone was not enough, as the different optimization objectives of the two branches could lead to a different emphases on the images and caused instability of the training process.
SA module, on the other hand, used the scale coefficient maps generated in SDC blocks to guide the dual branches to adopt the features from different receptive fields, which provided further flexibility and stability of the dual branch learning scheme.
Finally, we observed further performance gain by simultaneously incorporating both the FA and the SA modules.}

We acknowledge the limitation of this work.
In the experiments, we found that our method showed a certain level of trade-offs for the performance of the large lesions and small lesions.
Meanwhile, our method often achieved the best overall segmentation performance, yet it may not have the best performance for the small lesions and large lesions at the same time.
Therefore, we believe that further weighing the model performance of lesions at various scales will be a promising future work.
\textcolor{black}{In addition, the training efficiency of our method was decreased compared with that of the baselines, such as Unet. However, during testing, we used the LMSR path for segmentation and removed all other modules, and the inference efficiency is almost the same with the backbone.
}

\section{Conclusion}\label{conclusion}
In this work, we addressed the challenges of lesion segmentation raised by the low-resolution medical image and the diverse sizes of lesions.
First, our dual super-resolution framework could generate high-resolution images and segmentation masks from the low-resolution inputs.
Second, the proposed SDC blocks dynamically adjusted the network's receptive field, which largely improved the performance for the lesions of various sizes.
Further, we introduced the dual affinity learning modules to guide the LMSR branch to learn richer high-resolution information from the LISR branch, enhancing the robustness of the multi-task learning.
We have conducted extensive experiments on four challenging lesion segmentation datasets and demonstrated the effectiveness and generality of our proposed method on both 3D and 2D images.

% or


%\appendices

%\section{}

% \section*{Acknowledgment}




\bibliographystyle{IEEEtran}
\bibliography{reference}
\end{document}